\documentclass[aps,prb,twocolumn,notitlepage,floatfix,amsmath,amssymb,superscriptaddress]{revtex4-2}

\usepackage[T1]{fontenc}
\usepackage{bm}
\usepackage{booktabs}
\usepackage{graphicx}
\usepackage{mathtools}
\usepackage{microtype}
\usepackage[colorlinks=true,linkcolor=blue,citecolor=blue,urlcolor=blue]{hyperref}
\graphicspath{{figures/}}

\newcommand{\ii}{\mathrm{i}}

\newcommand{\mcL}{\mathcal{L}}
\newcommand{\mcC}{\mathcal{C}}
\newcommand{\mcO}{\mathcal{O}}
\newcommand{\mcT}{\mathcal{T}}

\begin{document}

\title{Complete kinematic null for local kinetic dissipation in a sixfold-driven electron fluid}

\author{P. Shubham Parashar}
\email{psparash@ucsd.edu}
\affiliation{Department of Physics, University of California, San Diego, La Jolla, California 92093, USA}

\date{August 10, 2026}

\begin{abstract}
We identify a complete kinematic null in a two-dimensional electron fluid driven
by a sixfold boundary pattern.  In the $U_3^\pm$ channel of $D_{12}$, device
symmetry excludes both the vector representation $R_1$ and the full
first-gradient tensor $R_1\otimes R_1$.  Hence at the symmetry-fixed center
$\bm j(\bm 0)=0$ and $\partial_i j_j(\bm 0)=0$, so every local quadratic
dissipative form built from the modeled charge/momentum field through first
gradient order vanishes independently of the constitutive coefficients.  In an
$O(2)$-isotropic angular-harmonic kinetic model the first allowed local sector is
$m=3$.  For $\nu q^2/\gamma_3\ll1$, its heating is set by the previously derived
coefficient $\kappa_4=v_F^4/(16\gamma_2^2\gamma_3)$.  An explicit incompressible
Stokes disk realizes a nonzero center signal, and a finite-moment kinetic
boundary-value solution approaches the corresponding local benchmark.  Within
the local, fixed-current momentum-conserving Stokes regime, the normalized
magnetic response can then be fitted for effective $\gamma_3$ and $\gamma_2$.
Thus the sixfold drive creates a measurement point where ordinary local
charge/momentum hydrodynamic dissipation is absent while a kinetic mode remains
finite.
\end{abstract}

\maketitle

\section{Introduction}
\label{sec:intro}

Electron hydrodynamics is now accessible through a combination of transport
and spatially resolved probes.  Nonlocal resistance, superballistic constriction
flow, Poiseuille profiles, vortices, and Hall-viscous response have been developed
theoretically and observed experimentally in high-mobility two-dimensional
conductors~\cite{Torre2015,LevitovFalkovich2016,Bandurin2016,KrishnaKumar2017,Moll2016,Sulpizio2019,Ku2020,Vool2021,AharonSteinberg2022,Scaffidi2017,Berdyugin2019,Gurzhi1968,LucasFong2018}.
A complementary experimental channel is dissipation itself: nanoscale
thermometry resolves where energy is deposited, and thermal transport
measurements have exposed viscous electronic
heating~\cite{Halbertal2016,Talanov2024}.  Local heating is therefore an
observable in its own right, not merely a diagnostic of the flow.

Cook and Lucas showed that device symmetry can select which dissipative channel
is reported by heating at a symmetry-fixed point~\cite{CookLucas2021}.  In their
square \(D_8\) example, irrep-pure boundary currents resolve the bulk,
rotational, two shear, and Ohmic channels, with one local hydrodynamic component
retained in each measurement channel.  They also analyzed nonhydrodynamic
corrections when the selected hydrodynamic contribution is absent, emphasizing
their different size dependence.  Thus both the representation-theoretic
selection method and symmetry-filtered kinetic residuals are prior ingredients of
the present construction.

In parallel, the kinetic content of these fluids has become experimentally
accessible.  In the tomographic regime, two-dimensional Fermi liquids can possess
anomalously long-lived odd angular harmonics, so that the \(m=3\) relaxation
rate \(\gamma_3\) is parametrically smaller than the even-sector rate
\(\gamma_2\)~\cite{Ledwith2019,HofmannGran2023,NilssonGranHofmann2025}.  Several
established probes already access these rates individually.  High-order cyclotron
resonance resolves the lifetime of the \(m\)th harmonic through the \(m\omega_c\)
structure of the absorption, and has been used to compare the second- and
third-order resonances directly~\cite{Moiseenko2025}.  A weak magnetic field
suppresses tomographic transport once the cyclotron radius reaches the odd-mode
mean free path, and channel and Corbino magnetotransport separate the even and
odd
scales~\cite{Rostami2025,BenShacharHofmannChannel2025,BenShacharHofmannCorbino2026,Starkov2026Corbino,MakiHofmann2026}.
Multiterminal geometry can also constrain momentum-relaxing,
momentum-conserving, and odd-sector rates, including \(\gamma_3\), from current
partition without spatial imaging~\cite{FarrellLucas2026}.

A different measurement condition is possible: symmetry may eliminate the
entire local charge/momentum kinematic data through first gradient order, rather
than selecting one hydrodynamic component, while leaving a higher kinetic sector
available.  In the Cook--Lucas viscometer, a non-vector viscosity drive makes the
center velocity vanish but retains the velocity-gradient component being
measured.  Magnetic and multiterminal probes instead infer rates from bulk or
contact-integrated responses.  Here we ask for the stronger fixed-point condition
in which both the vector field and its complete first spatial gradient vanish.

For a sixfold drive, the representation \(U_3^\pm\) of \(D_{12}\) lies outside the vector-plus-rank-two content of the one-component charge/momentum theory. It therefore enforces
\[
\mathbf{j}(0)=0,
\qquad
\partial_i j_j(0)=0,
\]
so every local quadratic dissipative form built from these variables vanishes independently of its constitutive coefficients. We refer to this simultaneous vanishing of the current and its complete first spatial gradient at the symmetry-fixed point as a \emph{complete kinematic null}. Within the declared isotropic kinetic model the first
sector that survives is \(m=3\); its magnitude is governed by a fourth-order
closure coefficient derived previously~\cite{ParasharClosure2026}, which is
imported here rather than re-derived.  An explicit disk solution shows the
surviving amplitude is constructively nonzero and supplies a model-specific
benchmark, a finite-moment kinetic calculation confirms it survives a microscopic
boundary condition, and a fixed-current magnetic sweep is then evaluated inside
the null.

\section{The complete kinematic null}
\label{sec:null}

\subsection{Setting}

We consider a linear, steady, two-dimensional kinetic electron fluid with an
\(O(2)\)-isotropic circular Fermi surface.  The distribution is expanded in
angular harmonics,
\begin{equation}
 f(\bm x,\vartheta)=\sum_m f_m(\bm x)e^{\ii m\vartheta},
 \label{eq:harmonics}
\end{equation}
streaming couples \(m\to m\pm1\), and the collision operator is taken diagonal in
angular momentum with scalar rates \(\gamma_m\), the \(m=0,1\) sectors being
conserved or only weakly relaxed.  The device and its boundary drive have the
dihedral symmetry \(H=D_{2M}\subset O(2)\), with \(D_{2M}\) of order \(2M\)
describing an \(M\)-fold device; the symmetry is imposed by the contacts and does
not require a crystal of the same point group.  The observation point
\(\bm x_\star\) is the symmetry-fixed center.  Nontrivial two-dimensional irreps
of \(O(2)\) are written \(R_m\), and \(U_k^\pm\) denote one-dimensional irreps
with reflection parity \(\pm\).

\subsection{Fixed-point representation content}

For the local charge/momentum dissipation considered here, the kinematic data
through first gradient order are the current vector,
\begin{equation}
 \bm j\in R_1,
\end{equation}
and its complete first spatial gradient is the rank-two tensor
\((\nabla\bm j)_{ij}\equiv\partial_i j_j\), which decomposes as
\begin{equation}
 \partial_ij_j\in R_1\otimes R_1
 =U_0^+\oplus U_0^-\oplus R_2
 \equiv\mcT_2,
 \label{eq:T2}
\end{equation}
whose three pieces are the trace (compression), the antisymmetric part
(vorticity), and the traceless symmetric part (shear).  Evaluation at a symmetry-fixed point is an \(H\)-equivariant map from
the drive representation to the local tensor representation.  By Schur's lemma,
the fixed-point value of a local quantity in representation \(R\) can be nonzero
for a drive in an irrep \(S_0\) only if
\(S_0\subset R\!\downarrow_H\)~\cite{CookLucas2021}.

Restricting to \(D_{12}\) (\(M=6\)), the harmonic index folds modulo \(M\), and
the reducible endpoint \(m=M/2=3\) splits:
\begin{equation}
 \mcT_2\!\downarrow_{D_{12}}=U_0^+\oplus U_0^-\oplus R_2,
 \qquad
 \bm j\in R_1,
 \label{eq:D12hydro}
\end{equation}
\begin{equation}
 R_3\!\downarrow_{D_{12}}=U_3^+\oplus U_3^-.
 \label{eq:R3branch}
\end{equation}
The device group \(D_{12}\) has the irreps
\(U_0^\pm,\,R_1,\,R_2,\,U_3^\pm\).  Comparing with Eq.~\eqref{eq:D12hydro}, the
hydrodynamic content uses \(U_0^\pm\), \(R_2\) and \(R_1\), and leaves
\begin{equation}
 \boxed{U_3^\pm\not\subset
 \left(\mcT_2\oplus R_1\right)\!\downarrow_{D_{12}}.}
 \label{eq:leftover}
\end{equation}
Within the dihedral family defined above, \(M=6\) is the smallest fold for which
a device irrep can lie outside \(\mcT_2\oplus R_1\); the corresponding leftover
is precisely \(U_3^\pm\).  The general branching statement is given in the
Supplemental Material.

\subsection{The null and its coefficient independence}

Let the imposed drive transform in \(U_3^\pm\).  In a linear equivariant
boundary-value problem, the response remains in the driven irrep sector.  By Eq.~\eqref{eq:leftover} neither \(\bm j\) nor
\(\partial_ij_j\) contains \(U_3^\pm\), so at the symmetry-fixed center
\begin{equation}
 \boxed{\;\bm j(\bm 0)=0,
 \qquad
 \partial_ij_j(\bm 0)=0.\;}
 \label{eq:completenull}
\end{equation}
For the linearized one-component charge/momentum fluid \(\bm v\propto\bm j\) at
the hydrodynamic order of interest, so the same statement holds for \(\bm v\) and
\(\partial_iv_j\).

Equation~\eqref{eq:completenull} is a statement about the fields themselves, not
about a transport coefficient.  Any local dissipative density that is a quadratic
form in the charge/momentum vector and its complete first gradient,
\begin{equation}
 Q_{\rm loc}
 =\Lambda^{(1)}_{ij}\,j_i j_j
 +\Lambda^{(2)}_{ij,kl}\,\partial_i j_j\,\partial_k j_l
 +\Lambda^{(3)}_{i,kl}\,j_i\,\partial_k j_l,
 \label{eq:quadform}
\end{equation}
therefore vanishes at \(\bm x_\star\) for arbitrary constitutive coefficient
tensors \(\Lambda^{(n)}\) within this sector.  This includes momentum-relaxing (Ohmic) dissipation, bulk and
shear viscous dissipation, rotational/vorticity dissipation where present, and
every symmetry-allowed cross term among these variables.  No assumption about the numerical values or tensor structure of the
transport coefficients within this declared local sector is used.

At fixed nonzero perpendicular magnetic field the reflections are broken and the
device symmetry reduces to the rotation subgroup \(C_6\).  Writing \(\chi_r\) for
the character with rotation eigenvalue \(e^{\ii r\pi/3}\),
\begin{equation}
 U_0^\pm\to\chi_0,
 \quad
 R_1\to\chi_1\oplus\chi_5,
 \quad
 R_2\to\chi_2\oplus\chi_4,
 \label{eq:C6hydro}
\end{equation}
whereas \(U_3^\pm\to\chi_3\).  The two reflection parities merge, but the selected
character remains absent from the vector and first-gradient hydrodynamic content,
so Eq.~\eqref{eq:completenull} survives on rotational symmetry alone.

\subsection{Relation to symmetry-selected viscometry}

The structural comparison with symmetry-selected viscometry is simple.  In the
Cook--Lucas square \(D_8\) decomposition, the rank-two plus vector hydrodynamic
content exhausts the available device irreps, with each channel associated with a
dissipative coefficient~\cite{CookLucas2021}.  A drive in any of the four non-vector viscosity irreps of that decomposition
does give \(\bm j(\bm 0)=0\), but the gradient component belonging to the
selected irrep is exactly the quantity that remains nonzero -- it is what makes
the measurement of that viscosity possible.  The remaining \(R_1\) channel is
the Ohmic vector channel, for which the center current itself is the selected
hydrodynamic variable.  Their channels are
coefficient filters, and their treatment of the residual heating in a suppressed
channel is likewise prior work.

\begin{figure*}[t]
 \centering
 \includegraphics[width=0.98\textwidth]{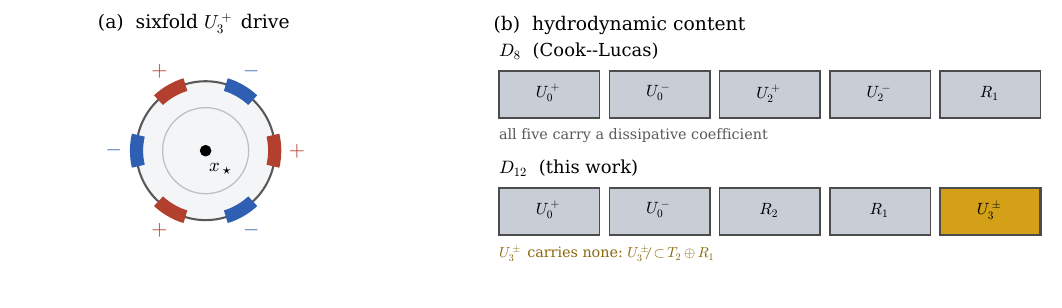}
 \caption{\label{fig:concept}\textbf{A device channel with no hydrodynamic
 content.}
 \textbf{(a)} Six equal normal contacts with alternating source and drain
 realize an exact \(U_3^+\) pattern; \(\bm x_\star\) is the symmetry-fixed
 center.
 \textbf{(b)} Hydrodynamic content of the device irreps.  In the Cook--Lucas
 \(D_8\) worked decomposition every device irrep carries one dissipative
 coefficient~\cite{CookLucas2021}.  Under \(D_{12}\) the same
 rank-two-plus-vector content uses only \(U_0^\pm\), \(R_2\) and \(R_1\), leaving
 \(U_3^\pm\) with no hydrodynamic coefficient, Eq.~\eqref{eq:leftover}.
}
\end{figure*}

The \(D_{12}\) restriction differs in one structural respect: the hydrodynamic
content does not exhaust the device irreps, and the leftover representation
\(U_3^\pm\) of Eq.~\eqref{eq:leftover} carries no hydrodynamic coefficient at
all.  Driving it therefore removes the fields rather than selecting a
coefficient.  Table~\ref{tab:compare} summarizes the two cases, and
Fig.~\ref{fig:concept} shows the same bookkeeping graphically.

\begin{table*}[t]
\caption{\label{tab:compare}Comparison of the two situations at the
symmetry-fixed center.  Both use the same representation-theoretic method;
they differ in whether the hydrodynamic content exhausts the device irreps.}
\centering
\begin{ruledtabular}
\begin{tabular}{lll}
 & \(D_8\) worked decomposition & \(D_{12}\), \(U_3^\pm\) drive \\
\hline
irreps used by \(\mcT_2\oplus R_1\)
 & all five
 & \(U_0^\pm,R_2,R_1\) only \\
leftover irrep
 & none
 & \(U_3^\pm\) \\
current at center
 & \(0\) in four viscosity irreps; \(R_1\) is the Ohmic vector channel
 & \(\bm j(\bm 0)=0\) \\
first gradient
 & selected component \(\neq0\) in viscosity channels
 & \(\partial_ij_j(\bm 0)=0\) \\
what is isolated
 & one dissipative coefficient
 & no local hydrodynamic form \\
\end{tabular}
\end{ruledtabular}
\end{table*}

Two limits of the statement should be kept in view from the outset.
Equation~\eqref{eq:completenull} constrains only the modeled local
charge/momentum sector through first gradient order.  It does not by itself null
dissipation carried by independent slow fields that were not part of the
representation analysis -- heat and energy modes, imbalance, valley or
spin-valley modes, multicomponent relative currents, phonon channels, or heat
generated elsewhere and transported to the observation point.  It is also a
statement of linear response; nonlinear products of the drive can regenerate
scalar channels.  We return to these in Sec.~\ref{sec:discussion}.

\section{First surviving kinetic sector}
\label{sec:kinetic}

\subsection{Which harmonic survives}

Equation~\eqref{eq:completenull} removes the hydrodynamic response but says
nothing about what remains.  Within the declared isotropic angular-harmonic
model, a sector \(m\) can carry a nonzero amplitude at the \(D_{12}\)-fixed
center only if its restriction contains the selected irrep.  By
Eq.~\eqref{eq:R3branch} the retained sectors \(m=0,1,2\) do not, while \(m=3\)
does.  Hence
\begin{equation}
 \boxed{m_\star=3,}
 \label{eq:mstar}
\end{equation}
and the leading local signal in this channel is kinetic rather than
hydrodynamic.

\subsection{Imported fourth-order closure input}

The magnitude of that surviving amplitude is fixed by results derived
previously~\cite{ParasharClosure2026}, which we state and use here rather than
re-derive.  Since streaming shifts the harmonic index by one, the shortest
excursion that leaves the current sector, reaches \(m=3\), and returns is
\begin{equation}
 1\rightarrow2\rightarrow3\rightarrow2\rightarrow1,
 \label{eq:path}
\end{equation}
whose four streaming vertices make the leading closure correction fourth order in
gradients~\cite{ParasharClosure2026}.  With
\(\partial_\pm=\partial_x\pm\ii\partial_y\), slaving the discarded harmonics to
the current at leading gradient order gives~\cite{ParasharClosure2026}
\begin{equation}
 \begin{aligned}
 f_{\pm2}&=-\frac{v_F}{2\gamma_2}\partial_{\mp}f_{\pm1}
 +\mcO(\nabla^3),\\
 f_{\pm3}&=\frac{v_F^2}{4\gamma_2\gamma_3}\partial_{\mp}^2f_{\pm1}
 +\mcO(\nabla^4).
 \end{aligned}
 \label{eq:f3local}
\end{equation}
Streaming flips harmonic parity at every vertex, so any path from \(m=1\) to
\(m=3\) contains an even number of vertices and the remainder in
Eq.~\eqref{eq:f3local} is \(\mcO(\nabla^4)\), not \(\mcO(\nabla^3)\).  The
associated fourth-order closure coefficient, together with the kinematic
viscosity, is~\cite{ParasharClosure2026}
\begin{equation}
 \boxed{
 \kappa_4=\frac{v_F^4}{16\gamma_2^2\gamma_3}=\frac{\nu^2}{\gamma_3},
 \qquad
 \nu=\frac{v_F^2}{4\gamma_2}.}
 \label{eq:kappa4}
\end{equation}
Equations~\eqref{eq:path}--\eqref{eq:kappa4}, the locality parameter below, and
the finite-field structure of Sec.~\ref{sec:field} are imported from
Ref.~\onlinecite{ParasharClosure2026}.  The distinct step here is the complete-null
condition under which this prior kinetic coefficient becomes the leading local
contribution.

The local collisional dissipation carried by the surviving sector is
\(\gamma_3(|f_{+3}|^2+|f_{-3}|^2)\), so at the fixed point
\begin{equation}
 \boxed{
 Q_{U_3}(\bm x_\star)
 =\kappa_4\left[
 |\partial_-^2f_{+1}|^2+|\partial_+^2f_{-1}|^2
 \right]_{\bm x_\star}}
 +\mcO(\nabla^6).
 \label{eq:Qlocal}
\end{equation}
Because the amplitude expansion advances in even powers, the first correction to
Eq.~\eqref{eq:Qlocal} in the bulk constitutive expansion is \(\mcO(\nabla^6)\);
there is no \(\mcO(\nabla^5)\) term.  This bulk statement should not be confused
with the finite-device contact-layer correction discussed in
Sec.~\ref{sec:finiteM}.

\subsection{Locality condition}

Eliminating \(m=3\) locally is controlled by the ratio of the fourth-order term
to the viscous one~\cite{ParasharClosure2026},
\begin{equation}
 \Xi_3(q)=\frac{\nu q^2}{\gamma_3}
 =\frac{v_F^2q^2}{4\gamma_2\gamma_3}
 =\frac{\ell_2\ell_3q^2}{4},
 \qquad
 \ell_m=\frac{v_F}{\gamma_m}.
 \label{eq:Xi3}
\end{equation}
Writing \(\Xi_3=(q\xi_3)^2\) defines the length
\begin{equation}
 \boxed{\xi_3=\sqrt{\frac{\nu}{\gamma_3}}=\tfrac12\sqrt{\ell_2\ell_3},}
 \label{eq:xi3}
\end{equation}
so that for a mode of characteristic wave number \(q=c/w\) with \(c=\mcO(1)\) the
local description requires
\begin{equation}
 \boxed{w\gg c\,\xi_3=\frac{c}{2}\sqrt{\ell_2\ell_3}},
 \qquad c=\mcO(1),
 \label{eq:localitycond}
\end{equation}
This is stronger than the condition \(w\gg\mcO(1)\,\ell_2\) that makes the stress
sector hydrodynamic, and the two differ precisely in the regime of interest: when
\(\gamma_3\ll\gamma_2\) one has \(\ell_3\gg\ell_2\) and
\(\xi_3=\tfrac12\ell_2\sqrt{\gamma_2/\gamma_3}\gg\ell_2\).  There is therefore a
parametrically allowed window \(\ell_2\ll w\ll\sqrt{\ell_2\ell_3}\) in which the
stress sector is hydrodynamic while \(m=3\) is not locally eliminable.  The
analytic expressions of this section and of Sec.~\ref{sec:field} apply outside
that window; the finite-moment solution of Sec.~\ref{sec:finiteM} makes no
gradient expansion, although we do not use it to make claims about the nonlocal
regime.  At finite field the exact control parameter is complex,
\(\Xi_\pm(q)=v_F^2q^2/(4\bar\lambda_{\pm2}\bar\lambda_{\pm3})\), and since
\(|\bar\lambda_{\pm m}|\ge\gamma_m\) for positive field-independent rates the
zero-field condition~\eqref{eq:localitycond} is sufficient across a sweep.

\section{Explicit disk realization}
\label{sec:disk}

Representation theory forbids the hydrodynamic response but does not guarantee
that the permitted kinetic amplitude is nonzero: a specially tuned drive could
still have vanishing overlap.  An explicit solution settles this.

Consider a disk of radius \(w\) driven by a smooth \(n=3\) normal current,
\begin{equation}
 j_r(w,\varphi)=j_0\cos3\varphi,
 \qquad
 j_\varphi(w,\varphi)=0,
 \label{eq:diskBC}
\end{equation}
which is charge conserving and transforms in \(U_3^+\); the pattern rotated by
\(\pi/6\) gives \(U_3^-\).  A discrete realization uses six equal normal contacts
at \(\varphi_j=j\pi/3\) with alternating currents \(I_j=(-1)^jI_0\), whose Fourier
content is confined to \(n\equiv3\!\pmod 6\) and is thus exactly irrep pure; the
first higher component \(n=9\) enters the center observable only at much higher
order (Supplemental Material, Sec.~\ref{sup:sec:comb}).

The regular incompressible Stokes stream function satisfying
Eq.~\eqref{eq:diskBC} exactly is
\begin{equation}
 \psi_+(r,\varphi)
 =j_0w\left[\frac56\left(\frac rw\right)^3
 -\frac12\left(\frac rw\right)^5\right]\sin3\varphi,
 \label{eq:psi}
\end{equation}
which obeys \(\nabla^4\psi_+=0\).  With \(j_r=r^{-1}\partial_\varphi\psi_+\) and
\(j_\varphi=-\partial_r\psi_+\), the center expansion is
\begin{equation}
 \begin{aligned}
 j_r&=\frac{5j_0}{2}\left(\frac rw\right)^2\cos3\varphi+\mcO(r^4),\\
 j_\varphi&=-\frac{5j_0}{2}\left(\frac rw\right)^2\sin3\varphi+\mcO(r^4).
 \end{aligned}
 \label{eq:centercurrent}
\end{equation}
so that the current vanishes quadratically and its complete first gradient
vanishes linearly,
\begin{equation}
 \bm j(\bm 0)=0,
 \qquad
 \left.\partial_ij_k\right|_{\bm 0}=0,
 \label{eq:diskchecks}
\end{equation}
in agreement with Eq.~\eqref{eq:completenull} and now verified componentwise
without recourse to representation theory.  Both the Ohmic density \(\propto|\bm j|^2\) and every local first-gradient
quadratic density therefore vanish at the center.

The surviving kinetic amplitude, by contrast, does not vanish.  Writing
\(z=x+\ii y\), the current near the center corresponds to
\(f_{+1}=A_3(z/w)^2+\mcO(r^4)\) with \(A_3\neq0\), and since
\(\partial_-=2\partial_z\),
\begin{equation}
 \left.\partial_-^2f_{+1}\right|_{\bm 0}=\frac{8A_3}{w^2}\neq0.
 \label{eq:nonzero}
\end{equation}
One further derivative annihilates the leading polynomial, so the shortest
streaming path consumes exactly the available degree.  The permitted signal is
thus constructively realized, not merely allowed.

Normalizing to the boundary current harmonic of one complex \(J=3\) block,
\(I_J=-\pi[f_1^{(3)}(w)+f_{-1}^{(3)}(w)]\), the regular no-slip Stokes disk gives
\(\partial_-^2f_1^{(3)}(0)=20I_J/(\pi w^3)\), and Eq.~\eqref{eq:Qlocal} yields
\begin{equation}
 \boxed{Q(\bm 0)=\frac{400}{\pi^2}\,
 \kappa_4\,\frac{|I_J|^2}{w^6}.}
 \label{eq:Chydro}
\end{equation}
The number \(400/\pi^2=40.5284734569\ldots\) is specific to this disk geometry,
this current normalization, and this hydrodynamic (no-slip) boundary condition;
it is not a universal coefficient, and it changes under a different normalization
or boundary law.  The full Stokes algebra and the general-\(n\) construction are
given in Supplemental Material, Sec.~\ref{sup:sec:disk}.

\begin{figure*}[t]
 \centering
 \includegraphics[width=0.92\textwidth]{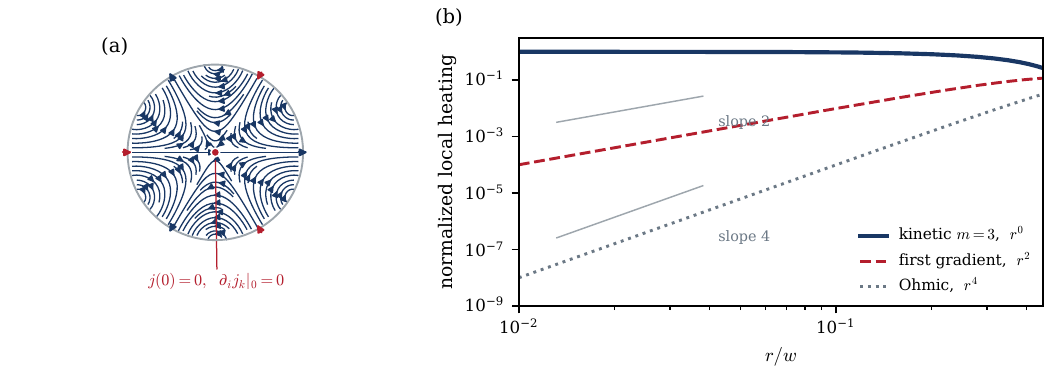}
 \caption{\label{fig:disk}\textbf{Explicit realization of the complete
 kinematic null.}
 \textbf{(a)} Streamlines of the exact biharmonic Stokes solution
 Eq.~\eqref{eq:psi} for the sixfold drive.  At the symmetry-fixed center both
 \(\bm j(\bm 0)=0\) and \(\partial_ij_k|_{\bm 0}=0\),
 Eq.~\eqref{eq:diskchecks}.
 \textbf{(b)} Local heating along a radius, each contribution normalized to its
 own leading coefficient.  The \(m=3\) kinetic contribution is finite at the
 center, whereas the first-gradient and Ohmic contributions vanish as \(r^2\)
 and \(r^4\); the surviving amplitude is therefore constructively nonzero,
 Eq.~\eqref{eq:nonzero}.}
\end{figure*}

\section{Magnetic response inside the null}
\label{sec:field}

The rate dependence of the surviving amplitude can be varied in situ with a
perpendicular magnetic field.  We emphasize at the outset that magnetic access to
individual angular-harmonic relaxation rates, including \(\gamma_3\) and its
comparison with \(\gamma_2\), is established
work~\cite{Moiseenko2025,Rostami2025,BenShacharHofmannChannel2025,BenShacharHofmannCorbino2026,Starkov2026Corbino,MakiHofmann2026},
as is geometric inference of the same rates from a designed
device~\cite{FarrellLucas2026}.  The finite-field kinetic structure used below --
the helicity denominators and the associated harmonic field scales -- is likewise
taken from Ref.~\onlinecite{ParasharClosure2026}.  What the present construction
adds is only the condition under which the response is read: the rate-sensitive
field dependence is evaluated at a point where, by
Eq.~\eqref{eq:completenull}, the local charge/momentum hydrodynamic response is
absent by symmetry rather than separated numerically or suppressed by a small
coefficient.

At dc the harmonic denominators become~\cite{ParasharClosure2026}
\begin{equation}
 \begin{aligned}
 \bar\lambda_{\pm m}&=\gamma_m\pm\ii m\omega_c,\\
 f_{\pm3}(B)&=\frac{v_F^2}{4\bar\lambda_{\pm2}\bar\lambda_{\pm3}}
 \partial_\mp^2f_{\pm1}+\mcO(\nabla^4).
 \end{aligned}
 \label{eq:f3B}
\end{equation}
We assume a semiclassical field range in which Landau quantization is negligible
and the rates \(\gamma_m\) do not vary appreciably across the sweep.  For constant coefficients in the simply connected incompressible Stokes disk,
the Lorentz and Hall-viscous forces are curl-free and can be absorbed into the
electrochemical pressure.  The fixed-current flow profile is therefore unchanged
in the momentum-conserving Stokes limit, so the common derivative amplitude and
current normalization cancel between finite and zero field, and
with \(|\bar\lambda_{\pm2}|^2=\gamma_2^2+4\omega_c^2\),
\(|\bar\lambda_{\pm3}|^2=\gamma_3^2+9\omega_c^2\),
\begin{equation}
 \boxed{
 \frac{Q_{U_3}(B)}{Q_{U_3}(0)}
 =\frac{\gamma_2^2\gamma_3^2}
 {(\gamma_2^2+4\omega_c^2)(\gamma_3^2+9\omega_c^2)}.}
 \label{eq:fieldratio}
\end{equation}
Equation~\eqref{eq:fieldratio} is prefactor-free within the local, fixed-current,
momentum-conserving Stokes regime of the scalar-rate model, and only there; it is
not a geometry-independent statement in general.  Geometry re-enters through
finite momentum relaxation, finite-optical-thickness boundary layers, a different
boundary law, nonlocal kinetics, or additional slow fields.

The two factors carry the characteristic scales~\cite{ParasharClosure2026}
\begin{equation}
 \omega_3=\frac{\gamma_3}{3},
 \qquad
 \omega_2=\frac{\gamma_2}{2},
 \qquad
 \frac{\gamma_3}{\gamma_2}=\frac32\frac{\omega_3}{\omega_2},
 \label{eq:knees}
\end{equation}
which are well separated when \(\gamma_3\ll\gamma_2\), although
Eq.~\eqref{eq:fieldratio} itself does not require that separation.  Within the
scalar-rate model a calibrated sweep can accordingly be fitted to infer the
effective rates \(\gamma_3\) and \(\gamma_2\); this is a model-dependent
inference, not a model-independent determination of a microscopic collision law.

\section{Finite-moment kinetic disk}
\label{sec:finiteM}

The preceding sections use a hydrodynamic flow profile and a local gradient
expansion.  To confirm that the surviving amplitude persists under a microscopic
boundary condition, we solve the truncated kinetic hierarchy in the disk without
making the gradient expansion.  Writing one conserved-\(J\) block with
\(\alpha=\vartheta-\varphi\), the radial ladder identities reduce the regular
solution to a finite Bessel superposition~\cite{ParasharClosure2026},
\begin{equation}
 f_m^{(J)}(r)=\sum_{a=1}^{m_{\max}}c_a\,u_m^{(a)}\,J_{J-m}(\kappa_ar),
 \label{eq:besselmodes}
\end{equation}
with exactly \(m_{\max}\) independent regular radial modes at cutoff \(|m|\le m_{\max}\),
matching the \(m_{\max}\) incoming characteristics of the truncated streaming operator.
Because \(J_{J-m}(\kappa r)\sim r^{|J-m|}\), only \(m=J\) survives at the center,
so the null-channel observable is a single amplitude per block.  We close the
problem with one declared microscopic boundary model: an ideal isotropic
reservoir imposed on the incoming characteristic subspace of the
Fourier--Galerkin truncation, \(W_-^\dagger f^{(J)}(w)=W_-^\dagger e_0\).  The
construction, mode counting and boundary matching are given in Supplemental
Material, Secs.~\ref{sup:sec:radial} and \ref{sup:sec:boundary}.

Using \(\gamma_3=0.1\gamma_2\), \(\gamma_{m\ge4}=\gamma_2\) with positive
regulators \(\gamma_0=\gamma_1=10^{-8}v_F/w\) for the conserved sectors, we
define the dimensionless center coefficient
\(\mathcal C_{m_{\max}}=w^6Q_{m_{\max}}^{(I)}(0)/\kappa_4\).  The calculation establishes three points.  First, the center signal remains
nonzero in the microscopic boundary-value problem.  Second, the reported
observables are converged in moment order for the declared spectrum: at \(w/\ell_2=100\) the coefficient differs between
\(m_{\max}=6\) and \(m_{\max}=10\) by \(9.5\times10^{-4}\) relative, and the field curve
differs between \(m_{\max}=6\) and \(m_{\max}=12\) by at most \(4.67\times10^{-5}\) over the
sampled field points.  Third, it approaches the local asymptotic benchmarks: at
\(m_{\max}=6\),
\begin{equation}
 \mathcal C_6=
 \begin{cases}
 37.3379,& w/\ell_2=100,\\
 39.7967,& w/\ell_2=400,\\
 40.3507,& w/\ell_2=1600,
 \end{cases}
 \label{eq:C6}
\end{equation}
the last lying \(0.44\%\) below \(400/\pi^2\), while the maximum deviation of the
normalized field curve from Eq.~\eqref{eq:fieldratio} falls from
\(2.97\times10^{-2}\) at \(w/\ell_2=50\) to \(7.35\times10^{-4}\) at
\(w/\ell_2=1600\), decreasing approximately as \((w/\ell_2)^{-1.09}\).  The remaining displacement at finite optical thickness persists after moment
convergence and is associated with the kinetic contact layer; it is a finite-device
correction distinct from the bulk constitutive remainder of
Sec.~\ref{sec:kinetic}.  Convergence tables, the regulator audit and the
regulator checks are in Supplemental Material, Sec.~\ref{sup:sec:convergence}.

The reported runs lie firmly in the local regime: with \(\gamma_3/\gamma_2=0.1\)
and \(q=1/w\), \(\Xi_3=2.5(w/\ell_2)^{-2}\), so even the smallest optical
thickness used has \(\Xi_3\simeq4\times10^{-3}\).  We therefore do not use these
data to make claims about the nonlocal regime.  The absolute coefficient at
finite optical thickness remains boundary-model dependent: reflecting gaps,
Maxwell specularity, partially transmitting contacts and matrix-valued radial
collision blocks are outside this calculation.

\begin{figure}[t]
 \centering
 \includegraphics[width=0.99\columnwidth]{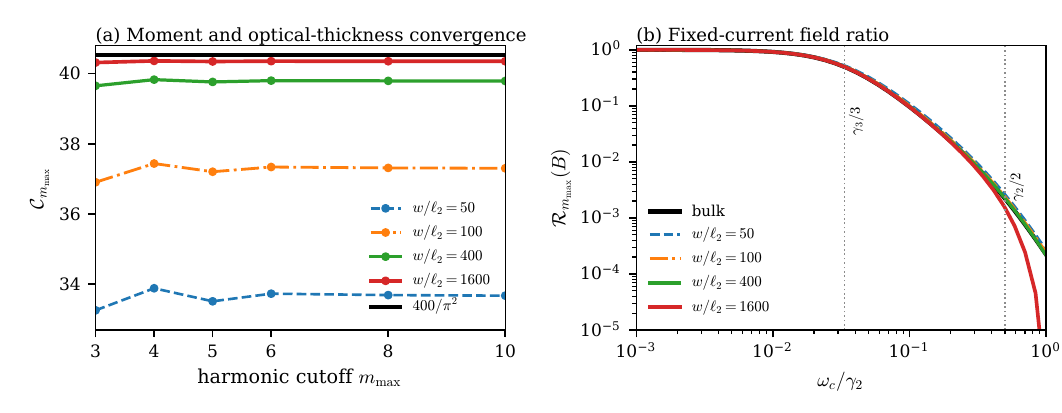}
 \caption{\label{fig:finiteM}\textbf{Finite-moment kinetic disk.}
 \textbf{(a)} Dimensionless center coefficient \(\mathcal C_{m_{\max}}\) versus harmonic cutoff
 \(m_{\max}\) for several optical thicknesses; the line is the model-specific Stokes
 value \(400/\pi^2\) of Eq.~\eqref{eq:Chydro}.  Moment convergence is rapid for
 the declared spectrum, and the residual displacement is a finite contact-layer
 correction that decreases with \(w/\ell_2\).
 \textbf{(b)} Fixed-current normalized field response of the complete \(m_{\max}=6\)
 disk approaching Eq.~\eqref{eq:fieldratio} (black).  Vertical guides mark
 \(\gamma_3/3\) and \(\gamma_2/2\).}
\end{figure}

\section{Discussion and limitations}
\label{sec:discussion}

Two ingredients should be kept conceptually separate.  Symmetry-selected
viscometry already shows how device irreps can suppress hydrodynamic channels and
leave kinetic residuals, while the $m=3$ closure coefficient and its magnetic
structure are known from prior kinetic work~\cite{CookLucas2021,ParasharClosure2026}.
The additional condition studied here is kinematic rather than constitutive:
$U_3$ is absent from all local vector and first-gradient charge/momentum content,
so the corresponding hydrodynamic variables themselves vanish at the center.
This stronger condition is insensitive to how the allowed first-gradient
coefficients are partitioned or cross-coupled within the modeled sector: once
$\bm j$ and $\nabla\bm j$ vanish, those coefficients have no local kinematic
argument on which to act.  What survives is then fixed by the kinetic hierarchy;
in the present $O(2)$ scalar-rate model the first allowed sector is $m=3$.

That separation also clarifies what the null does not protect against.  The
calculated $Q(\bm r)$ is local collisional dissipation, whereas thermometry
measures a temperature field shaped by thermal diffusion, phonon coupling, probe
convolution, and heat generated elsewhere.  Likewise, the theorem concerns only
the modeled local charge/momentum variables through first gradient order;
independent heat, imbalance, valley or spin-valley, multicomponent, phonon, and
contact channels lie outside it.  Contact asymmetry or miscentering can admix
lower irreps and generate a background.  The distinct center expansions of the
kinetic, first-gradient, and Ohmic contributions make a line scan through the
center useful as an internal diagnostic, but not a substitute for symmetric
calibration.  Nonlinear response can also regenerate hydrodynamic scalar
channels.

The local constitutive expressions, including Eqs.~\eqref{eq:Qlocal} and
\eqref{eq:fieldratio}, require the locality condition~\eqref{eq:localitycond}.
For a long-lived odd harmonic this is more restrictive than ordinary stress
hydrodynamics, leaving a parametric window in which the $m=2$ sector is
hydrodynamic while $m=3$ is not locally eliminable; that nonlocal regime is not
addressed by the present asymptotic formulas.  Equation~\eqref{eq:fieldratio}
further assumes negligible momentum relaxation on the device scale, so that the
fixed-current profile is the momentum-conserving Stokes solution rather than a
field-dependent Gurzhi flow.

The size and temperature dependences are therefore best viewed as consistency
tests.  At fixed current and geometry the leading center heating scales as
$w^{-6}$ with $Q\propto1/(\gamma_2^2\gamma_3)$.  Common Fermi-liquid rates then
give $Q\propto T^{-6}$, while an anomalously long-lived $m=3$ harmonic with
$\gamma_3\propto T^4$~\cite{Ledwith2019,HofmannGran2023,NilssonGranHofmann2025}
gives $Q\propto T^{-8}$.  These powers depend on the collision spectrum and are
not symmetry-protected.  The symmetry statement is instead that, within the
modeled local charge/momentum sector, the ordinary hydrodynamic contribution is
removed kinematically before those kinetic scalings are examined.

\section{Conclusion}
\label{sec:conclusion}

We have shown that a sixfold $D_{12}$ boundary drive admits a $U_3$ channel
outside the complete vector-plus-rank-two hydrodynamic content of an isotropic
electron fluid.  The resulting fixed point satisfies $\bm j(\bm 0)=0$ and
$\partial_i j_j(\bm 0)=0$, while an $m=3$ kinetic amplitude remains allowed.  In
the local regime this makes the known fourth-order kinetic coefficient the
leading modeled center dissipation; the Stokes disk and finite-moment calculation
verify that the allowed signal is nonzero and approaches the local benchmark.
A fixed-current magnetic sweep then provides a model-dependent rate diagnostic
inside the same null.  The essential consequence is therefore a measurement
condition in which conventional local charge/momentum hydrodynamic heating is
removed before the kinetic response is read.

\bibliographystyle{apsrev4-2}
\bibliography{references}

\clearpage
\onecolumngrid
\begin{center}
{\large\bfseries Supplemental Material for\\[2pt]
``Complete kinematic null for local kinetic dissipation in a sixfold-driven electron fluid''}\\[8pt]
P. Shubham Parashar\\
Department of Physics, University of California, San Diego, La Jolla, California 92093, USA
\end{center}
\vspace{0.5em}

\renewcommand{\thesection}{S\arabic{section}}
\renewcommand{\thesubsection}{S\arabic{section}.\arabic{subsection}}
\renewcommand{\theequation}{S\arabic{equation}}
\renewcommand{\thefigure}{S\arabic{figure}}
\renewcommand{\thetable}{S\arabic{table}}
\renewcommand{\theHsection}{S\arabic{section}}
\renewcommand{\theHsubsection}{S\arabic{section}.\arabic{subsection}}
\renewcommand{\theHequation}{S\arabic{equation}}
\renewcommand{\theHfigure}{S\arabic{figure}}
\renewcommand{\theHtable}{S\arabic{table}}
\setcounter{section}{0}
\setcounter{equation}{0}
\setcounter{figure}{0}
\setcounter{table}{0}

\section{Branching and the leftover-irrep criterion}
\label{sup:sec:branching}

Restriction of an \(O(2)\) irrep to the device group \(D_{2M}\) folds the
harmonic index modulo \(M\)~\cite{CookLucas2021},
\begin{equation}
 R_m\!\downarrow_{D_{2M}}=R_{f_M(m)},
 \qquad
 f_M(m)=\min\left(m\bmod M,\;M-(m\bmod M)\right),
 \label{sup:eq:branching}
\end{equation}
with reducible endpoints
\begin{equation}
 R_0=U_0^+\oplus U_0^-,
 \qquad
 R_{M/2}=U_{M/2}^+\oplus U_{M/2}^-\quad(M\ \text{even}).
 \label{sup:eq:endpoints}
\end{equation}
Fixed-point heating in a measurement irrep \(S_0\) can receive a contribution
from a fluid irrep \(R\) only if \(S_0\subset R\!\downarrow_{D_{2M}}\).

For the local charge/momentum sector considered here, the kinematic data through first gradient order are the vector \(\bm j\in R_1\) and its complete first gradient
\(\partial_ij_j\in R_1\otimes R_1=U_0^+\oplus U_0^-\oplus R_2\equiv\mcT_2\).  We
therefore call a device irrep \(S_0\) a \emph{leftover} irrep when
\begin{equation}
 \boxed{S_0\not\subset\mcT_2\!\downarrow_{D_{2M}}\;\oplus\;R_1\!\downarrow_{D_{2M}},}
 \label{sup:eq:leftovercriterion}
\end{equation}
in which case a drive in \(S_0\) forces both \(\bm j\) and \(\partial_ij_j\) to
vanish at the fixed point.  Applying Eq.~\eqref{sup:eq:branching}:

\begin{itemize}
\item \(M=4\) (\(D_8\)): \(R_1\!\downarrow=R_1\), \(R_2\!\downarrow=U_2^+\oplus U_2^-\),
 so \(\mcT_2\oplus R_1=U_0^+\oplus U_0^-\oplus U_2^+\oplus U_2^-\oplus R_1\).
 These are all five irreps of \(D_8\); there is no leftover irrep, consistent
 with the correspondence between the five \(D_8\) channels and the bulk,
 rotational, two shear and Ohmic dissipative coefficients~\cite{CookLucas2021}.
\item \(M=6\) (\(D_{12}\)): \(R_2\!\downarrow=R_2\), so
 \(\mcT_2\oplus R_1=U_0^+\oplus U_0^-\oplus R_2\oplus R_1\), while \(D_{12}\)
 also possesses \(U_3^\pm\).  These satisfy
 Eq.~\eqref{sup:eq:leftovercriterion} and are reached first by \(R_3\), since
 \(R_3\!\downarrow_{D_{12}}=U_3^+\oplus U_3^-\).
\end{itemize}

More generally a leftover irrep exists whenever the device group possesses an
irrep of folded index \(f_M(m)\ge3\), i.e. for \(M\ge6\); the \(D_{12}\) case is
the smallest such device and the one used in the main text.

\section{Equivariance of exact moment elimination}
\label{sup:sec:schur}

The selection rule is not an artifact of truncation.  Let \(U_h\) represent
\(h\in H\), let \(P\) project onto the retained moment sector and \(Q=1-P\), and
assume \([\mcL,U_h]=0\) and \([P,U_h]=0\).  The exact retained-sector operator is
the Schur complement
\begin{equation}
 \mcL_{\rm eff}=P\mcL P-P\mcL Q\,(Q\mcL Q)^{-1}Q\mcL P.
 \label{sup:eq:schur}
\end{equation}
Every factor commutes with \(U_h\), hence \([\mcL_{\rm eff},U_h]=0\): exact
elimination of the discarded moments cannot mix inequivalent device irreps.  The
Schur-complement construction itself is standard and, in the circular kinetic
setting used here, is taken from Ref.~\onlinecite{ParasharClosure2026}.

\section{Device-channel table}
\label{sup:sec:table}

Table~\ref{sup:tab:orders} records, for the \(O(2)\)-isotropic angular hierarchy
restricted to a device subgroup, the lowest discarded harmonic \(m_\star>2\)
whose restriction contains the measurement irrep, and the corresponding gradient
order \(p_{\rm cl}=2(m_\star-1)\) obtained by counting streaming vertices.  These
entries describe an isotropic fluid decomposed into device channels; they are
\emph{not} a statement about the kinetics of an intrinsically anisotropic
crystal, whose velocity operator and collision kernel are not of the form assumed
here.  The \(m_\star=4\) entry reproduces the \(w^{-8}\) size scaling of the
Cook--Lucas square viscometer~\cite{CookLucas2021} and is retained only as a
consistency check on the vertex counting.

\begin{table}[h]
\caption{\label{sup:tab:orders}Device-channel content of the \(O(2)\)-isotropic
kinetic hierarchy under dihedral device subgroups.  ``Isotropic device-channel
content'' names the local hydrodynamic object carrying the channel in the
isotropic fluid.}
\centering
\begin{ruledtabular}
\begin{tabular}{c c c l c c}
\(M\) & device & irrep & isotropic device-channel content & \(m_\star\) & \(p_{\rm cl}\)\\
\hline
3 & \(D_6\)    & \(U_0^+\) & compressional (bulk) & 3 & 4\\
3 & \(D_6\)    & \(U_0^-\) & vorticity (rotational) & 3 & 4\\
3 & \(D_6\)    & \(R_1\)   & vector/Ohmic (repeated) & 4 & 6\\
\hline
4 & \(D_8\)    & \(U_0^+\) & compressional (bulk) & 4 & 6\\
4 & \(D_8\)    & \(U_0^-\) & vorticity (rotational) & 4 & 6\\
4 & \(D_8\)    & \(R_1\)   & vector/Ohmic & 3 & 4\\
\hline
6 & \(D_{12}\) & \(U_0^+\) & compressional (bulk) & 6 & 10\\
6 & \(D_{12}\) & \(U_0^-\) & vorticity (rotational) & 6 & 10\\
6 & \(D_{12}\) & \(R_2\)   & shear & 4 & 6\\
6 & \(D_{12}\) & \(R_1\)   & vector/Ohmic & 5 & 8\\
6 & \(D_{12}\) & \(U_3^\pm\) & \textbf{none} (leftover irrep) & 3 & 4\\
\end{tabular}
\end{ruledtabular}
\end{table}

The last row is the channel used in the main text: it is the only entry whose
device-channel content is empty, which is precisely the leftover-irrep condition
of Eq.~\eqref{sup:eq:leftovercriterion}.

\section{Imported fourth-order closure input}
\label{sup:sec:prior}

For self-containment we reproduce the elements of
Ref.~\onlinecite{ParasharClosure2026} used in the main text.  They are results of
that work and are not derived anew here.

With \(\partial_\pm=\partial_x\pm\ii\partial_y\), the streaming operator is
\(\bm v\cdot\nabla=(v_F/2)(e^{\ii\vartheta}\partial_-+e^{-\ii\vartheta}\partial_+)\),
so projecting the static kinetic equation onto harmonic \(k\) gives, for scalar
rates and at dc,
\begin{equation}
 \frac{v_F}{2}\left(\partial_-f_{k-1}+\partial_+f_{k+1}\right)
 +\bar\lambda_kf_k=0,
 \qquad
 \bar\lambda_k=\gamma_{|k|}+\ii k\omega_c.
 \label{sup:eq:chain}
\end{equation}
Slaving the discarded harmonics to \(f_{\pm1}\) at leading gradient order yields
\begin{equation}
 f_{\pm2}=-\frac{v_F}{2\bar\lambda_{\pm2}}\partial_\mp f_{\pm1},
 \qquad
 f_{\pm3}=\frac{v_F^2}{4\bar\lambda_{\pm2}\bar\lambda_{\pm3}}\partial_\mp^2f_{\pm1},
 \label{sup:eq:slaving}
\end{equation}
so that the shortest excursion leaving and returning to the current sector is
\(1\to2\to3\to2\to1\), with four streaming vertices and hence a fourth-order
closure correction.  At zero field this gives
\begin{equation}
 \kappa_4=\frac{v_F^4}{16\gamma_2^2\gamma_3}=\frac{\nu^2}{\gamma_3},
 \qquad
 \nu=\frac{v_F^2}{4\gamma_2},
 \label{sup:eq:kappa4supp}
\end{equation}
and the gradient expansion is controlled by
\(\Xi_3=\nu q^2/\gamma_3=\ell_2\ell_3q^2/4\).

\paragraph*{Remainder order.}
Streaming flips harmonic parity at every vertex, so any path from \(m=1\) to
\(m=3\) contains an even number of vertices and the local constitutive expansion
has the structure \(f_3=f_3^{(2)}+f_3^{(4)}+\cdots\).  The correct sharp
statement is therefore
\begin{equation}
 f_3=f_3^{(2)}+\mcO(\nabla^4),
 \label{sup:eq:remainder}
\end{equation}
and consequently
\begin{equation}
 \gamma_3|f_3|^2=\gamma_3|f_3^{(2)}|^2
 +2\gamma_3\,\mathrm{Re}\!\left[(f_3^{(2)})^*f_3^{(4)}\right]+\cdots,
\end{equation}
so the leading heating is \(\mcO(\nabla^4)\) and the next \emph{bulk
constitutive} correction is \(\mcO(\nabla^6)\); there is no \(\mcO(\nabla^5)\)
term.  This bulk statement is distinct from the finite-device contact-layer
correction of Sec.~\ref{sup:sec:convergence}, which is a boundary effect and decays
approximately as \(\ell_2/w\).

\section{Analytic disk construction}
\label{sup:sec:disk}

For a boundary harmonic \(n\ge2\) with \(j_r(w,\varphi)=j_0\cos n\varphi\) and
\(j_\varphi(w,\varphi)=0\), the regular incompressible Stokes stream function is
\begin{equation}
 \psi_n(r,\varphi)=j_0w\left[\frac{n+2}{2n}\left(\frac rw\right)^n
 -\frac12\left(\frac rw\right)^{n+2}\right]\sin n\varphi,
 \label{sup:eq:psin}
\end{equation}
which satisfies \(\nabla^4\psi_n=0\) and both boundary conditions exactly.  With
\(z=x+\ii y\) and \(j_+=j_x+\ii j_y\), the center behavior is
\begin{equation}
 j_+=\frac{n+2}{2}j_0\left(\frac{\bar z}{w}\right)^{n-1}+\mcO(r^{n+1}),
 \label{sup:eq:jplus}
\end{equation}
so for a real distribution with \(j_+\propto f_{-1}=f_{+1}^*\) one has
\(f_{+1}=A_n(z/w)^{n-1}+\mcO(r^{n+1})\) with \(A_n\neq0\).  Since
\(\partial_-=2\partial_z\),
\begin{equation}
 \left.\partial_-^{\,n-1}f_{+1}\right|_{\bm 0}
 =\frac{2^{n-1}(n-1)!\,A_n}{w^{n-1}}\neq0,
 \label{sup:eq:sharpderiv}
\end{equation}
while one further derivative annihilates the leading polynomial: the shortest
streaming path consumes exactly the available degree.  The subleading
\(r^{n+2}\) term contributes a current of degree \(n+1\) and does not affect
Eq.~\eqref{sup:eq:sharpderiv}.

For \(n=3\), Eq.~\eqref{sup:eq:psin} reduces to the stream function of the main text,
and \(\partial_-^2f_{+1}(0)=8A_3/w^2\neq0\).  A direct symbolic check confirms
\(\nabla^4\psi_+=0\), \(\bm j(\bm 0)=0\), and the vanishing of all four
components of \(\partial_ij_k\) at the center.  Normalizing to
\(I_J=-\pi[f_1^{(3)}(w)+f_{-1}^{(3)}(w)]\), the regular no-slip Stokes disk gives
\(f_1^{(3)}(r)=5I_Jz^2/(2\pi w^3)+\mcO(r^4)\), hence
\(\partial_-^2f_1^{(3)}(0)=20I_J/(\pi w^3)\) and
\begin{equation}
 \mathcal C_{\rm hydro}=\frac{400}{\pi^2}=40.5284734569\ldots
 \label{sup:eq:Chydrosupp}
\end{equation}
for this geometry, this current normalization and this no-slip hydrodynamic
boundary condition.  It is not a universal constant and changes under a different
normalization or boundary law.

\section{Finite-moment radial reduction}
\label{sup:sec:radial}

The circular reduction and radial ladder machinery used here follow
Ref.~\onlinecite{ParasharClosure2026}.  With \(\alpha=\vartheta-\varphi\) and
\(f(r,\varphi,\vartheta)=e^{\ii J\varphi}\sum_{m=-m_{\max}}^{m_{\max}}f_m^{(J)}(r)e^{\ii m\alpha}\),
the static kinetic equation in one \(J\) block is
\begin{equation}
 \left[v_F\cos\alpha\,\partial_r
 +\frac{v_F\sin\alpha}{r}\left(\ii J-\partial_\alpha\right)
 +\omega_c\partial_\alpha+\mcC_{m_{\max}}\right]F_J(r,\alpha)=0,
 \label{sup:eq:Jblock}
\end{equation}
and projecting onto \(e^{\ii k\alpha}\) gives
\begin{equation}
 \frac{v_F}{2}\left[\mathcal A_{k-1}f^{(J)}_{k-1}+\mathcal B_{k+1}f^{(J)}_{k+1}\right]
 +\bar\lambda_kf^{(J)}_k=0,
 \qquad
 \mathcal A_mg=g'+\frac{J-m}{r}g,
 \quad
 \mathcal B_mg=g'-\frac{J-m}{r}g.
 \label{sup:eq:radialhier}
\end{equation}
The ladders act diagonally on Bessel functions of matched order,
\begin{equation}
 \mathcal A_m\left[J_{J-m}(\kappa r)\right]=+\kappa J_{J-(m+1)}(\kappa r),
 \qquad
 \mathcal B_m\left[J_{J-m}(\kappa r)\right]=-\kappa J_{J-(m-1)}(\kappa r),
 \label{sup:eq:besselladder}
\end{equation}
so \(f^{(J)}_m(r)=u_mJ_{J-m}(\kappa r)\) removes the radial dependence and leaves
the algebraic problem
\begin{equation}
 \bar\lambda_ku_k+\frac{v_F\kappa}{2}\left(u_{k-1}-u_{k+1}\right)=0.
 \label{sup:eq:dispersion}
\end{equation}
Writing this as \((\Lambda+tS)u=0\) with \(t=v_F\kappa/2\), the odd dimension
\(2m_{\max}+1\) makes \(S\) singular, so one root lies at infinity and \(2m_{\max}\) are
finite; the involution \(u_k\mapsto(-1)^ku_k\) maps \(t\mapsto-t\) and
\(J_\nu(-z)=(-1)^\nu J_\nu(z)\), so \(\pm\kappa_a\) generate the same radial
solution.  There are therefore exactly \(m_{\max}\) independent regular radial modes at
cutoff \(|m|\le m_{\max}\), matching the \(m_{\max}\) incoming characteristics at \(r=w\).
Since \(J_{J-m}(\kappa r)\sim r^{|J-m|}\),
\begin{equation}
 f^{(J)}_m(0)=0\quad\text{unless}\quad m=J,
 \label{sup:eq:centerselect}
\end{equation}
so the center observable is one amplitude per block, and for the \(U_3\) channel
\(Q(\bm 0)=\gamma_3[|f^{(3)}_3(0)|^2+|f^{(-3)}_{-3}(0)|^2]\).

\section{Boundary model}
\label{sup:sec:boundary}

A normal-current condition supplies only one boundary moment, whereas the
truncation admits \(m_{\max}\) incoming characteristics.  We close the problem with one
declared microscopic model: an ideal isotropic reservoir on the incoming
characteristic subspace of the Fourier--Galerkin truncation.  Let \(\mathsf C_{m_{\max}}\)
represent multiplication by \(\cos\alpha\), \((\mathsf C_{m_{\max}})_{m,m\pm1}=\tfrac12\),
with \(m_{\max}\) negative, \(m_{\max}\) positive and one zero eigenvalue, and let \(W_-\)
contain the negative-velocity eigenvectors.  Then
\begin{equation}
 W_-^\dagger f^{(J)}(w)=W_-^\dagger e_0,
 \label{sup:eq:inflow}
\end{equation}
with \(e_0\) the isotropic \(m=0\) reservoir mode, supplies exactly the \(m_{\max}\)
required conditions.  High-precision arithmetic is used because the hydrodynamic
and contact-layer modes separate exponentially at large optical thickness
\(K=w/\ell_2\).

This is one boundary model, not a universal electronic boundary law.  Reflecting
gaps, Maxwell specularity, partially transmitting contacts and matrix-valued
radial collision blocks are outside the present calculation and can change the
finite-\(K\) contact-layer correction and the absolute coefficient away from the
hydrodynamic limit.  They do not affect the exact radial reduction or the
symmetry selection of Sec.~\ref{sup:sec:branching}.

\section{Convergence and regulator checks}
\label{sup:sec:convergence}

The spectrum is \(\gamma_3=0.1\gamma_2\), \(\gamma_{m\ge4}=\gamma_2\), with
positive regulators \(\gamma_0=\gamma_1=10^{-8}v_F/w\) for the conserved sectors,
and \(\mathcal C_{m_{\max}}=w^6Q^{(I)}_{m_{\max}}(0)/\kappa_4\).

\emph{Moment convergence.}  At \(K=100\) the relative difference between
\(\mathcal C_6\) and \(\mathcal C_{10}\) is \(9.5\times10^{-4}\).  For the
normalized field curve at the same optical thickness, the maximum difference
between \(m_{\max}=6\) and \(m_{\max}=12\) over the sampled field points is
\(4.67\times10^{-5}\).  These are the two distinct comparisons available in the
production data: the coefficient is compared at \(m_{\max}=6\) versus \(m_{\max}=10\), and the
field curve at \(m_{\max}=6\) versus \(m_{\max}=12\).

\emph{Optical-thickness dependence.}  \(\mathcal C_6=37.3379,\,39.7967,\,40.3507\)
at \(K=100,400,1600\), the last \(0.44\%\) below \(400/\pi^2\).  The maximum
deviation of the \(m_{\max}=6\) field curve from the local bulk ratio decreases as in
Table~\ref{sup:tab:fielderr}, approximately as \(K^{-1.09}\).  The boundary residuals
in the high-precision solve are below \(10^{-80}\); the large double-precision
condition numbers reflect the exponential separation of bulk and contact-layer
modes, not failure of the matched solution.

\begin{table}[h]
\caption{\label{sup:tab:fielderr}Maximum field-ratio deviation of the complete
\(m_{\max}=6\) disk from the local bulk expression.}
\centering
\begin{ruledtabular}
\begin{tabular}{c c}
\(w/\ell_2\) & \(\max_B|\mathcal R_6-\mathcal R_{\rm bulk}|\)\\
\hline
50   & \(2.9733\times10^{-2}\)\\
100  & \(1.1938\times10^{-2}\)\\
200  & \(5.0921\times10^{-3}\)\\
400  & \(2.3278\times10^{-3}\)\\
800  & \(1.0903\times10^{-3}\)\\
1600 & \(7.3497\times10^{-4}\)\\
\end{tabular}
\end{ruledtabular}
\end{table}

\emph{Regulator audit.}  At \(m_{\max}=6\), \(K=100\), reducing the conserved-mode
regulator gives
\begin{align}
 \gamma_0=\gamma_1&=10^{-6},\;10^{-8},\;10^{-10},\;10^{-12},\\
 \mathcal C_{m_{\max}}&=37.337424,\;37.337865,\;37.337870,\;37.337870,
\end{align}
respectively, so the quoted value is regulator independent at the \(10^{-7}\)
relative level.

\emph{Regime of the production data.}  With \(\gamma_3/\gamma_2=0.1\) and
\(q=1/w\), \(\Xi_3=2.5K^{-2}\); the smallest optical thickness used has
\(\Xi_3\simeq4\times10^{-3}\).  All reported data therefore lie in the local
regime, and no claim is made about the nonlocal regime.

\begin{figure}[h]
 \centering
 \includegraphics[width=0.62\textwidth]{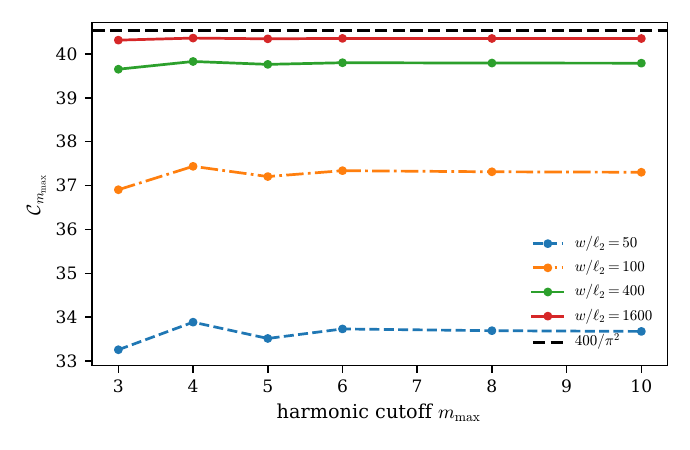}
 \caption{\label{sup:fig:mconv}Moment convergence of the center coefficient at
 several optical thicknesses; the dashed line is the model-specific value
 \(400/\pi^2\).}
\end{figure}

\begin{figure}[h]
 \centering
 \includegraphics[width=0.62\textwidth]{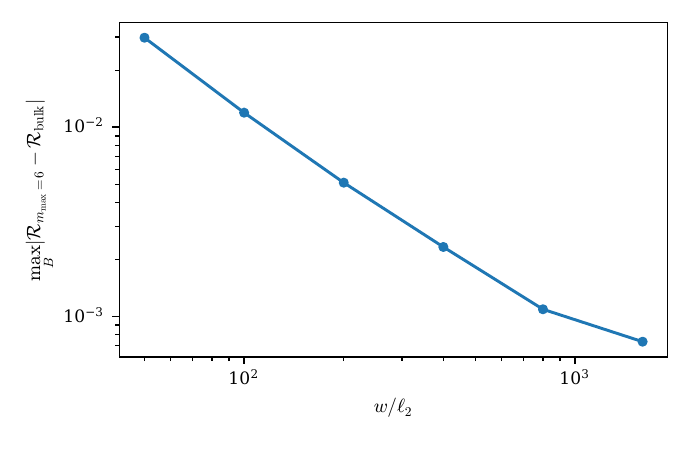}
 \caption{\label{sup:fig:fielderr}Contact-layer correction to the normalized field
 response: the maximum deviation of the complete \(m_{\max}=6\) disk from the local bulk
 expression decreases approximately inversely with \(w/\ell_2\).}
\end{figure}

\section{Contact-comb decomposition}
\label{sup:sec:comb}

The discrete six-contact drive is exactly irrep pure.  For the \(U_3^+\) comb
\begin{equation}
 J_+(\varphi)=\sum_{j=0}^{5}(-1)^j\delta_{2\pi}\!\left(\varphi-\frac{j\pi}{3}\right),
 \label{sup:eq:comb}
\end{equation}
the Fourier coefficient is the geometric sum
\begin{equation}
 \widetilde J_{+,n}=\sum_{j=0}^{5}(-1)^je^{-\ii nj\pi/3}
 =\begin{cases}6,& n\equiv3\pmod 6,\\ 0,&\text{otherwise},\end{cases}
 \label{sup:eq:combfourier}
\end{equation}
so that \(J_+(\varphi)=(6/\pi)\sum_{r\ge0}\cos[(6r+3)\varphi]\) and every spatial
harmonic belongs to the same \(U_3^+\) irrep.  The leading component \(n=3\)
controls the signal; the first higher component \(n=9\) requires eight spatial
derivatives to survive at the fixed point and therefore contributes only at order
\(\epsilon^{16}\) in the device-scale expansion.

\section{Finite-field structure}
\label{sup:sec:field}

The finite-field denominators and harmonic field scales are taken from
Ref.~\onlinecite{ParasharClosure2026}.  At dc,
\(\bar\lambda_{\pm m}=\gamma_m\pm\ii m\omega_c\), and Eq.~\eqref{sup:eq:slaving}
gives the field-dependent \(m=\pm3\) amplitude directly.  The dissipative rate of
that sector is \(\gamma_3\), so
\begin{equation}
 Q_{U_3}(B)=\frac{\gamma_3v_F^4}{16}
 \left[\frac{|\partial_-^2f_{+1}|^2}{|\bar\lambda_{+2}|^2|\bar\lambda_{+3}|^2}
 +\frac{|\partial_+^2f_{-1}|^2}{|\bar\lambda_{-2}|^2|\bar\lambda_{-3}|^2}\right]_{\bm x_\star}
 +\mcO(\nabla^6).
 \label{sup:eq:QB}
\end{equation}
In the momentum-conserving Stokes limit the fixed-current flow profile is field
independent, because the Lorentz and Hall-viscous forces are proportional to
\(\hat{\bm z}\times\bm j\) and \(\nabla^2(\hat{\bm z}\times\bm j)\), whose
two-dimensional curls are \(\nabla\cdot\bm j=0\) and
\(\nabla^2\nabla\cdot\bm j=0\) for incompressible flow, so both are absorbed into
the electrochemical pressure.  The derivative matrix element then cancels in the
ratio, and with \(|\bar\lambda_{\pm2}|^2=\gamma_2^2+4\omega_c^2\) and
\(|\bar\lambda_{\pm3}|^2=\gamma_3^2+9\omega_c^2\) one obtains the normalized
response quoted in the main text.  This cancellation holds within the local,
fixed-current, momentum-conserving Stokes regime of the scalar-rate model; finite
momentum relaxation makes the profile field dependent through the even viscosity
and reintroduces a geometry-dependent correction.

\begin{figure}[h]
 \centering
 \includegraphics[width=0.72\textwidth]{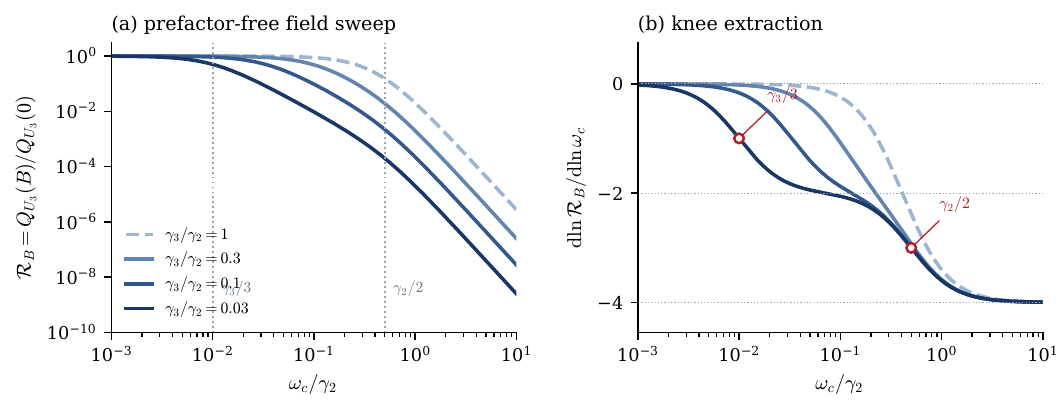}
 \caption{\label{sup:fig:sweep}Normalized fixed-current field response for several
 ratios \(\gamma_3/\gamma_2\), with the characteristic scales \(\gamma_3/3\) and
 \(\gamma_2/2\).  The curve is prefactor-free within the regime stated above and
 may be fitted for the effective rates.}
\end{figure}

\end{document}